\documentclass{aa}

\usepackage{graphicx}
\usepackage{txfonts}
\begin{document}

\title{Eruption of a multi-flux-rope system in solar active region 12673 leading to the two
       largest flares in Solar Cycle 24}

\author{Y. J. Hou \inst{1,2}
      \and J. Zhang \inst{1,2}
      \and T. Li \inst{1,2}
      \and S. H. Yang \inst{1,2}
      \and X. H. Li \inst{1,2}
      }

\institute{CAS Key Laboratory of Solar Activity, National Astronomical Observatories,
           Chinese Academy of Sciences, Beijing 100101, China\\
           \email{yijunhou@nao.cas.cn; zjun@nao.cas.cn}
           \and School of Astronomy and Space Science, University of Chinese Academy of Sciences, Beijing 100049, China
           }

\date{Received 22 December 2017; accepted 21 August 2018}


  \abstract
   {Solar active region (AR) 12673 in 2017 September produced the two largest flares in Solar Cycle 24: the X9.3
   flare on September 6 and the X8.2 flare on September 10.
   }
   {We attempt to investigate the evolutions of the two large flares and their associated complex magnetic
   system in detail.
   }
   {Combining observations from the \emph{Solar Dynamics Observatory} and results of nonlinear force-free
   field (NLFFF) modeling, we identify various magnetic structures in the AR core region and examine the evolution
   of these structures during the flares.
   }
   {Aided by the NLFFF modeling, we identify a double-decker flux rope configuration above the polarity
   inversion line (PIL) in the AR core region. The north ends of these two flux ropes were rooted in a negative-
   polarity magnetic patch, which began to move along the PIL and rotate anticlockwise before the X9.3 flare on
   September 6. The strong shearing motion and rotation contributed to the destabilization of the two magnetic
   flux ropes, of which the upper one subsequently erupted upward due to the kink-instability. Then another two
   sets of twisted loop bundles beside these ropes were disturbed and successively erupted within five minutes like
   a chain reaction. Similarly, multiple ejecta components were detected as consecutively erupting during the X8.2
   flare occurring in the same AR on September 10. We examine the evolution of the AR magnetic fields from September
   3 to 6 and find that five dipoles emerged successively at the east of the main sunspot. The interactions
   between these dipoles took place continuously, accompanied by magnetic flux cancellations and strong shearing
   motions.
   }
   {In AR 12673, significant flux emergence and successive interactions between the different emerging dipoles
   resulted in a complex magnetic system, accompanied by the formations of multiple flux ropes and twisted
   loop bundles. We propose that the eruptions of a multi-flux-rope system resulted in the two largest flares
   in Solar Cycle 24.
   }


\titlerunning{flux ropes, flares, and active region}
\authorrunning{Hou et al.}

\maketitle
%

\section{Introduction}

Solar flares are explosive phenomena on the Sun that can be observed from X-ray to radio wavelengths,
and that release dramatic free magnetic energy stored in the solar atmosphere via the process of magnetic
reconnection (Priest \& Forbes 2002; Schmieder et al. 2015). In previous studies, the accumulation
of free magnetic energy in the solar atmosphere is demonstrated to be mainly caused by three types
of mechanisms: (1) magnetic flux emergence or cancellation (Wang \& Shi 1993; Chen \& Shibata 2000;
Zhang et al. 2001; Sterling et al. 2010; Louis et al. 2015), (2) shearing motion (Wang et al. 1994;
Meunier \& Kosovichev 2003; Sun et al. 2012), (3) sunspot rotation (Brown et al. 2003; Zhang et al. 2007;
T{\"o}r{\"o}k et al. 2013). Although the energy accumulation has been investigated thoroughly, it is
difficult for us to comprehend the detailed process of violent energy release in various solar eruptions.
Because it is widely accepted that magnetic flux ropes play key roles in triggering eruptive events
(Amari et al. 2000; Fan 2005; Kliem et al. 2010; Liu et al. 2010; Green et al. 2011; Li et al. 2016;
Yan et al. 2017), we can understand these eruptive events such as solar flares and coronal mass ejections
(CMEs) through studying magnetic flux ropes.

\begin{figure*}
\centering
\includegraphics [width=0.96\textwidth]{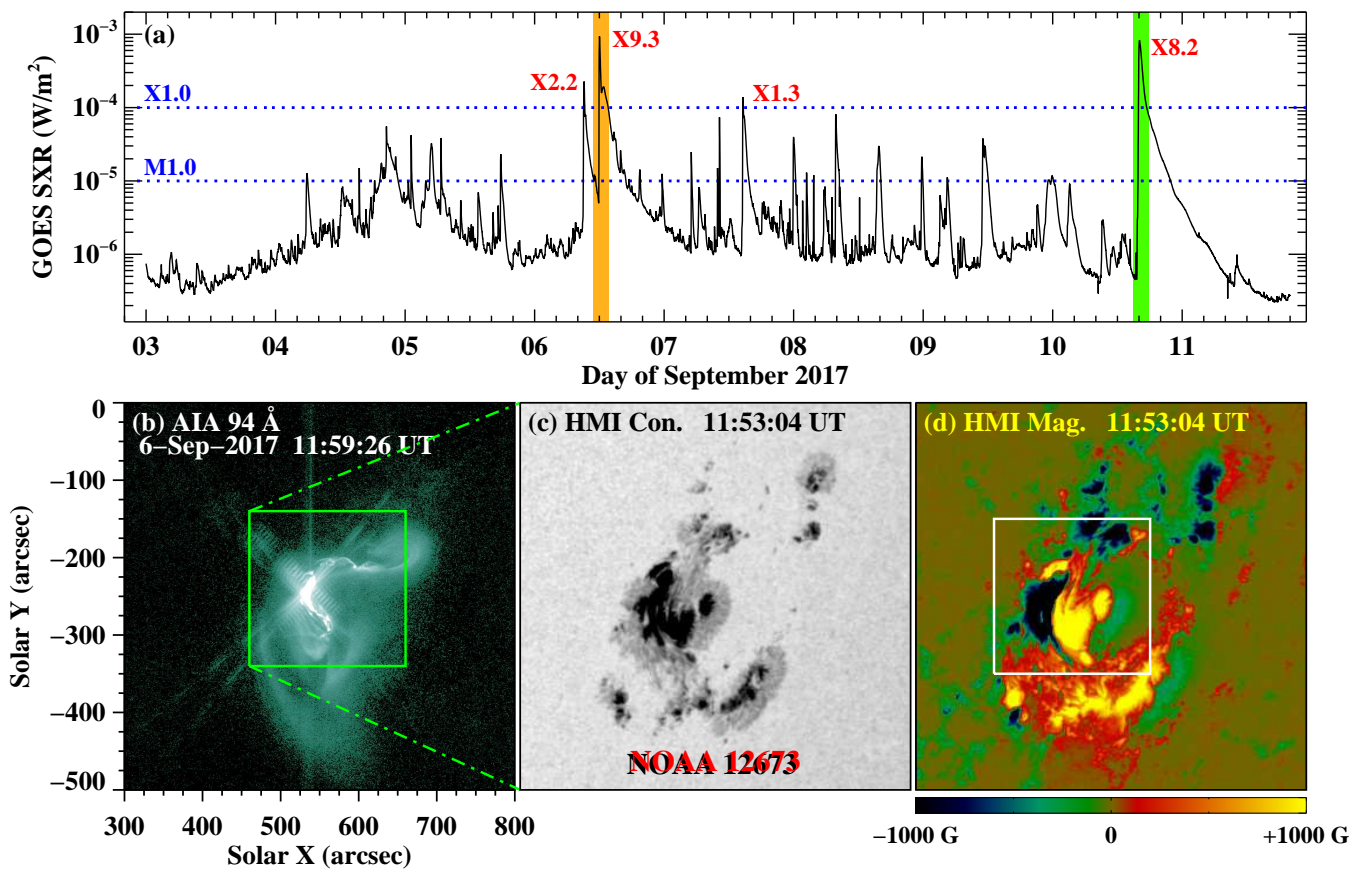}
\caption{
Flares produced by AR 12673.
Panel (a): GOES SXR 1-8 {\AA} flux variation from 2017 September 3 to September 11. Four X-class flares
took place during this period, of which the two largest ones reached up to X9.3 (orange region) and X8.2
(green region), respectively. The blue horizontal dotted lines mark the threshold levels of M1.0 and X1.0
flares. Panels (b)-(d): overview of the X9.3 flare in AR 12673 on September 6. The AIA 94 {\AA} image in
panel (b) shows this AR at the onset of the flare. HMI continuum intensitygram and LOS magnetogram in panels
(c) and (d) display the sunspots and underlying magnetic fields in the AR core region, whose field of view
(FOV) is outlined by the green square in panel (b).
}
\label{fig1}
\end{figure*}

A magnetic flux rope is a set of magnetic field lines winding around a central axis in classical eruptive
flare models and many CME observations. A huge amount of effort has been made in numerical simulations
of the formation and dynamic activity of flux ropes (Forbes \& Priest 1995; Aulanier et al. 2010).
Amari et al. (2000, 2003) simulated the evolution of a flux rope and proposed that a slow converging
motion of the footpoints of field lines toward the polarity inversion line (PIL) contributed to the
formation of a flux rope through magnetic reconnection. With high-resolution observations, the existence
of flux ropes in the solar atmosphere has also been recently evidenced (Guo et al. 2010, 2013;
Cheng et al. 2011; Yang et al. 2014; Kumar et al. 2017; Guglielmino et al. 2017; Wang et al. 2017a;
Yan et al. 2018a; Shen et al. 2018). Zhang et al. (2012) reported a flux rope observed as a hot
extreme ultraviolet (EUV) channel before and during the solar eruption and proposed that the instability of
this flux rope triggered the eruption. Li \& Zhang (2013a) investigated the successive eruptions of two flux
ropes during an M-class flare. Li \& Zhang (2013b) presented four homologous flux ropes, which were formed
successively at the same location in an active region (AR). These observations imply that flux ropes
may be ubiquitous on the Sun (Zhang et al. 2015; Hou et al. 2016). In the present work, a magnetic
flux rope is defined as a set of magnetic field lines winding around a central axis by more than one
full turn (Liu et al. 2016). Then aided by nonlinear force-free field (NLFFF) modeling and the calculation
of twist number, we can identify a magnetic flux rope without ambiguity.

Flux ropes are often related to various magnetohydrodynamic (MHD) instability processes, which eventually trigger
solar flares and CMEs (Alexander et al. 2006; Liu et al. 2007; Kumar \& Cho 2014). Kink MHD instability
is triggered by the azimuthal twist of magnetic tubes. Numerical simulations of the kink instability suggest
that if the twist of a flux rope exceeds a critical value, then this rope becomes unstable (Kliem et al. 2004).
The exact value of required twist depends on various factors such as loop geometry and overlying magnetic
fields (Hood \& Priest 1979; Baty 2001; Fan \& Gibson 2004; Leka et al. 2005). In addition, observations
of kink instability were reported recently by many authors (Srivastava et al. 2010; Wang et al. 2017b).

From 2017 September 4 to September 10, AR 12673 produced a total of 4 X-class flares, 27 M-class flares,
and a multitude of smaller ones (see the details in Yang et al. 2017). The X9.3 flare on September
6 is the largest flare in Solar Cycle 24 and has been reported in several works (Wang et al. 2018; Verma 2018;
Shen et al. 2018; Yan et al. 2018b; Jiang et al. 2018). In this work, we identify a double-decker flux rope
configuration above the PIL in the AR core region and detect successive eruptions of multiple flux ropes and
twisted loop bundles within five minutes before the peak of this large flare. A similar phenomenon was also
observed during the X8.2 flare on September 10. Here we investigate the evolutions of the two large flares
and the associated complex magnetic system in detail.

The remainder of this paper is structured as follows. Section 2 contains the observations and data analysis
taken in our study. The detailed process of the two flares and the evolution of the magnetic fields in the AR
core region are presented in Sect. 3. Finally, in Sect. 4 we conclude this work and discuss the results.

\begin{figure*}
\centering
\includegraphics [width=0.96\textwidth]{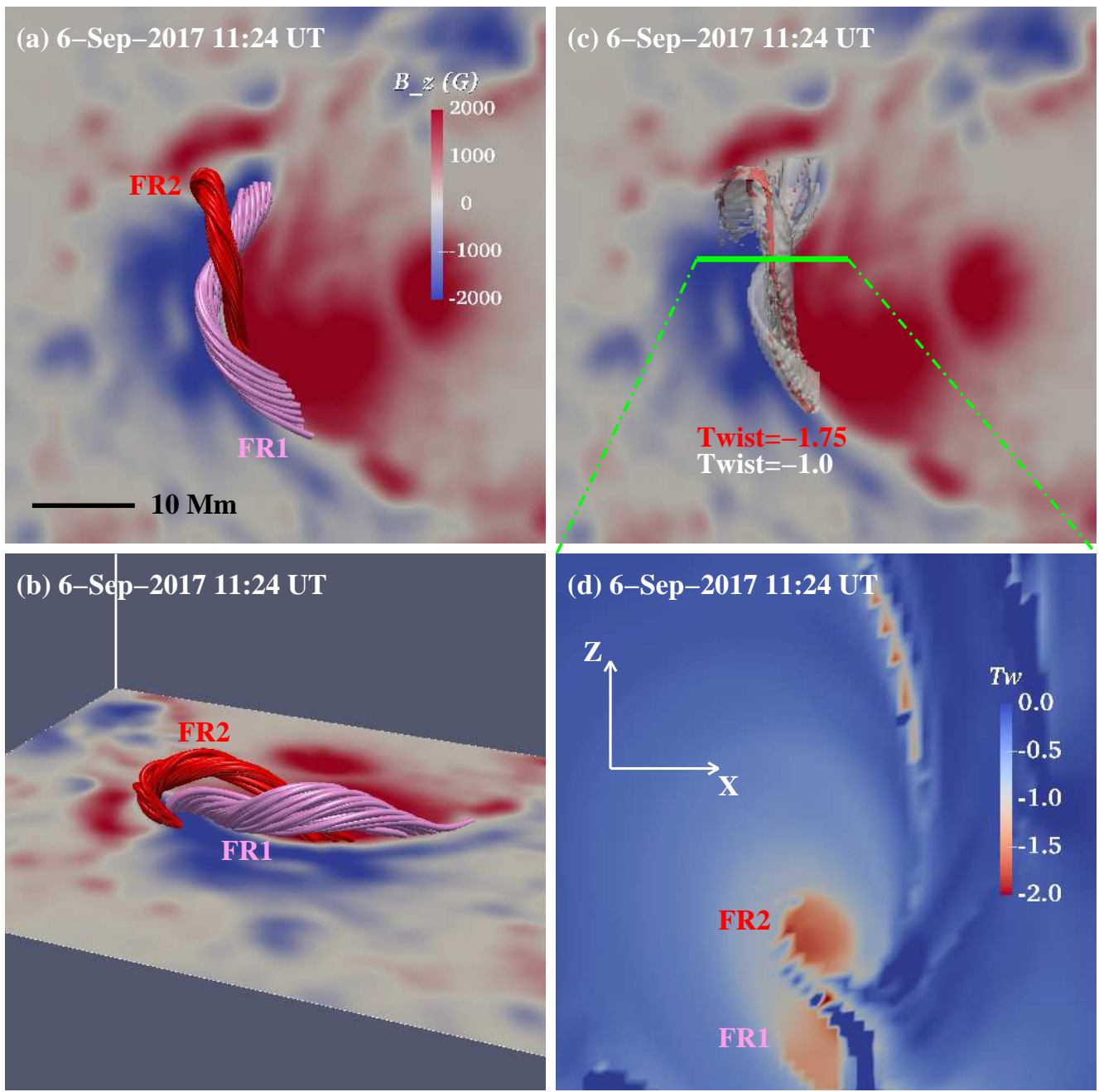}
\caption{
Double-decker flux rope configuration above the PIL in the AR core region revealed by NLFFF modeling at 11:24 UT
on 2017 September 6. Panels (a)-(b): top view and side view of two flux ropes (FR1 and FR2) composing the
double-decker configuration. The FOV of these panels is approximated by the white square in Fig. 1(d).
Panel (c): isosurfaces of twist number $T_{w}$=--1 (white) and $T_{w}$=--1.75 (red) viewed from the
same perspective as panel (a).
Panel (d): twist number distribution in the vertical (x-z) plane along the green cut labeled in panel (c).
}
\label{fig2}
\end{figure*}

\section{Observations and data analysis}

On 2017 September 6, an X9.3 flare took place in NOAA AR 12673, which was the largest flare in Solar Cycle 24.
About four days later, another X8.2 flare happened in the same AR when the AR rotated to the solar southwestern
limb on September 10. The Atmospheric Imaging Assembly (AIA; Lemen et al. 2012) on board the \emph{Solar
Dynamics Observatory} (\emph{SDO}; Pesnell et al. 2012) successively observes multilayered solar atmosphere in
ten (E)UV passbands with a cadence of (12)24 s and a spatial resolution of 1.{\arcsec}2. The
\emph{SDO}/Helioseismic and Magnetic Imager (HMI; Schou et al. 2012) provides one-arcsecond resolution
full-disk line-of-sight (LOS) magnetograms and intensitygrams every 45 s, and photospheric vector magnetograms
at a cadence of 720 s (Hoeksema et al. 2014). Here we employ the data of AIA 94 {\AA}, 171 {\AA}, 304 {\AA},
HMI LOS magnetograms, intensitygrams and the HMI data product called Space-weather HMI Active Region Patches
(SHARP; Bobra et al. 2014) for the investigation of the X9.3 flare on September 6. The AIA 131 {\AA} and 171
{\AA} images are used to study the X8.2 flare on September 10. Moreover, the observations from the
\emph{Geostationary Operational Environmental Satellite} (\emph{GOES}) are also used to present the
variation of soft X-ray (SXR) 1-8 {\AA} flux from September 3 to September 11.

The AIA and HMI observations used in the X9.3 flare are all derotated to the reference time of 12:02
UT on September 6, and the AIA data of the X8.2 flare are aligned to 16:06 UT on September 10. To
investigate the evolution of magnetic fields in AR 12673 before the onset of the X9.3 flare, we employ
the HMI LOS magnetograms and intensitygrams from September 3 to September 6 and derotate them to
a middle time of 00:00 UT on September 05. To determine the horizontal photospheric velocities, we
apply the method of differential affine velocity estimator (DAVE; Schuck 2006). The window size in
DAVE is set as 19 pixels, following the parameter given in Liu et al. (2013).

\begin{figure*}
\centering
\includegraphics [width=0.96\textwidth]{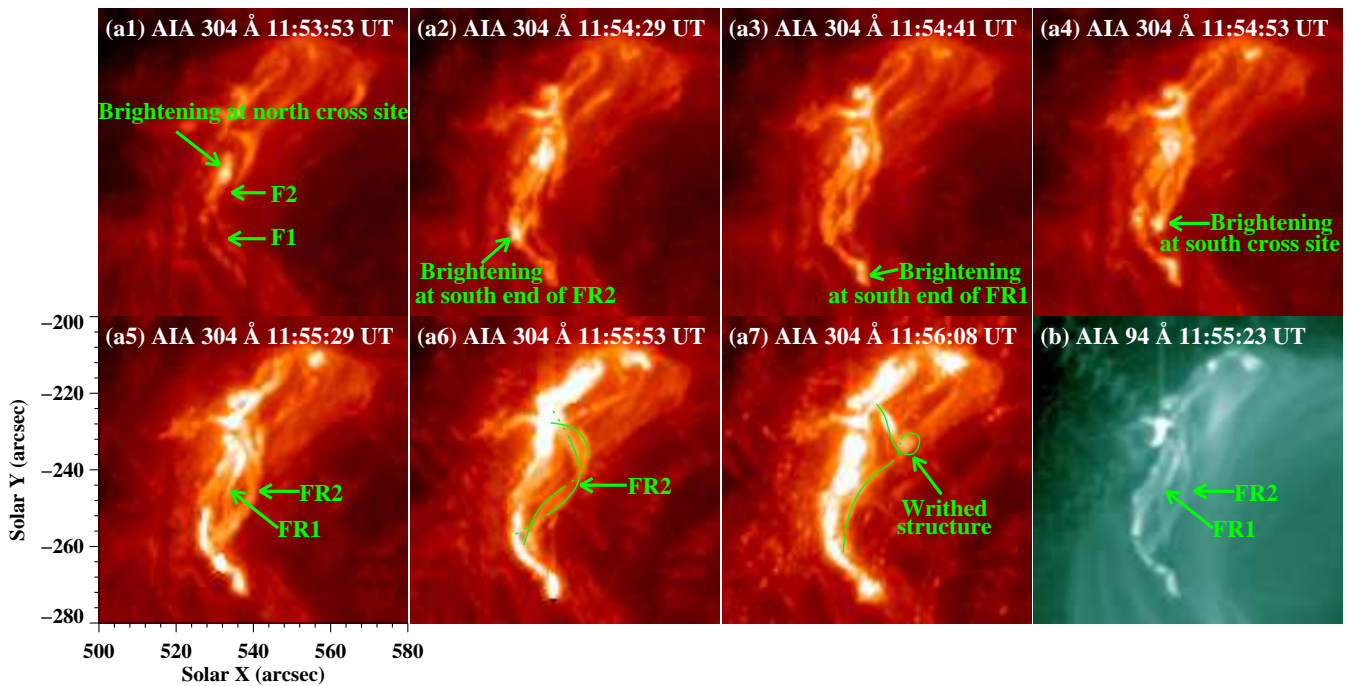}
\caption{
Eruption of the double-decker flux rope configuration.
Panels (a1)-(a7): sequence of extended AIA 304 {\AA} images showing the filament and eruption process of
the double-decker flux rope configuration. The two filaments and associated brightening before their eruption
are denoted by green arrows in panels (a1)-(a4). The green curves in panels (a6) and (a7) delineate the
twisted threads and the writhed structure of the upper flux rope (FR2).
Panel (b): corresponding 94 {\AA} image exhibiting the double-decker flux rope configuration in a
higher temperature wavelength. The FOV is outlined by the white square in Fig. 1(d).
An animation (1.mpg) of the 304 {\AA} and the 94 {\AA} images is available online.
}
\label{fig3}
\end{figure*}

In order to reconstruct the three-dimensional (3D) coronal structure in the target region, we utilize the
``weighted optimization" method to perform NLFFF extrapolation (Wiegelmann 2004; Wiegelmann et al. 2012)
based on the observed photospheric vector magnetic fields. To satisfy the force-free condition, the vector
magnetograms are preprocessed by a procedure developed by Wiegelmann et al. (2006) towards suitable
photospheric boundary conditions. The calculation is performed within a box of 512 $\times$ 448 $\times$ 256
uniform grid points (186 $\times$ 162 $\times$ 93 Mm$^{3}$), which covers nearly the entire AR. We further
calculate the twist number $T_{w}$ (Berger \& Prior 2006) of the extrapolated field using the code developed
by R. Liu and J. Chen (Liu et al. 2016).

\section{Results}
\subsection{Overview of AR 12673}

When AR 12673 approached the solar center on September 03, significant flux emergence commenced in this
region and persisted for the following several days (Sun \& Norton 2017). Before disappearing
completely in the west solar limb on September 11, AR 12673 produced a total of 4 X-class flares, 27
M-class flares, and numerous smaller ones. Figure 1(a) shows the \emph{GOES} SXR 1-8 {\AA} flux variation
from September 3 to September 11, and four X-class flares are denoted. The orange and green regions in
panel (a) respectively mark the two largest flares in Solar Cycle 24: the X9.3 flare on September 6 and
the X8.2 flare on September 10. In AIA 94 {\AA} channel, the X9.3 flare performed complex structures
accompanied by intense emission enhancement (panel (b)). This flare started around 11:53 UT and peaked at
12:02 UT when the AR was centered around S09W34. Panels (c) and (d) show the HMI continuum intensitygram
and LOS magnetogram of the AR at the onset of the flare.

\subsection{The X9.3 flare on September 6}

Based on the photospheric vector magnetograms observed by \emph{SDO}/HMI, we extrapolated the 3D structure
of the AR using NLFFF modeling at 11:24 UT on September 6, just before the occurrence of the X9.3 flare.
For visualizations of the magnetic field above the region of interest, we select a region from the NLFFF
extrapolation, with an FOV approximated by the white box in Fig. 1(d), to display in Fig. 2. Moreover, we
calculate the twist number $T_{w}$ of the reconstructed field and then we can obtain the photospheric twist
map or vertical twist map in the selected cutting plane. According to the photospheric twist map and the
definition of magnetic flux rope mentioned in Sect. 1, we plot the field lines across the photosphere where
the $\mid$$T_{w}$$\mid$ $\geqq$ 1.0 near the PIL in the AR core region. As a result, we
obtain two magnetic flux ropes with $T_{w}$ $\leqq$ --1.0, of which the upper one (FR2) is located right
above the lower one (FR1), forming a double-decker flux rope configuration (Liu et al. 2012). Figures 2(a)
and 2(b) show the top view and side view of the double-decker flux rope configuration, respectively. The
background is the photospheric vertical magnetic field (B$_z$). Panel (c) exhibits the isosurfaces of
$T_{w}$=--1 (white) and $T_{w}$=--1.75 (red) above the PIL from the top view. It is clear that the
isosurface of $T_{w}$=--1 roughly outlines the two flux ropes and the $\mid$$T_{w}$$\mid$ around the axes
of the two flux ropes is beyond 1.75. Along the green cut marked in panel (c), we make a twist map in
the vertical (x-z) plane and show it in panel (d). One can see that there are two regions with high negative
twist number in this vertical twist map, which correspond to the cross sections of the two flux ropes.

\begin{figure*}
\centering
\includegraphics [width=0.96\textwidth]{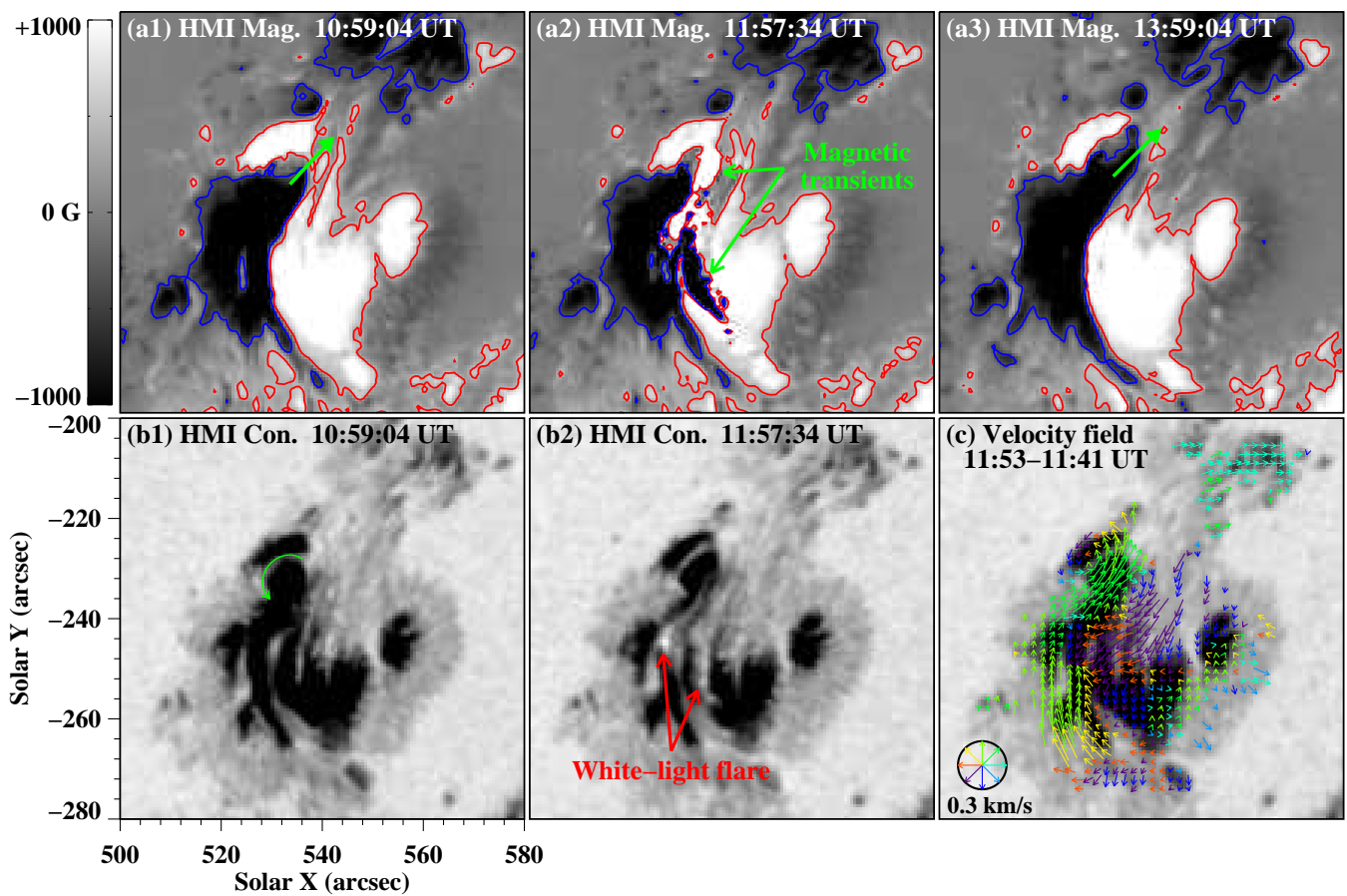}
\caption{
Magnetic evolution of the AR core region during the X9.3 flare.
Panels (a1)-(a3): HMI LOS magnetograms with contours of $\pm$400 G displaying the magnetic fields of the
core region before and after the flare peak at 12:02 UT. The green solid arrows in panels (a1) and (a3)
denote the direction of rapid displacement of the negative patch where the north ends of the double-decker
flux rope configuration are rooted. The green arrows in panel (a2) mark the magnetic transients caused by
the white-light flare.
Panels (b1)-(b2): corresponding HMI continuum intensitygrams. The green arrow in panel (b1) labels the
rotation of the negative patch. The red arrows in panel (b2) mark the signals of the white-light flare.
Panel (c): the horizontal photospheric velocity field (colored arrows) derived from the HMI continuum
intensitygrams computed by the DAVE method. In the lower-left corner, the radius of the circle corresponds to
a speed of 0.3 km s$^{-1}$, and the color of an arrow corresponds to its direction.
The FOV is the same as Fig. 3. An animation (2.mpg) of the HMI continuum intensitygrams and LOS magnetograms
is available online.
}
\label{fig4}
\end{figure*}

By examining the AIA 304 {\AA} observations, we detected two sets of filament threads located in the AR core
region before the occurrence of the X9.3 flare (see F1 and F2 in Fig. 3(a1) and the corresponding animation).
According to Fig. 2, we suggest that F1 corresponds to the upper flux rope FR1 and F2 corresponds to the lower
rope FR2. Around 11:53:53 UT, brightening appeared at the north cross site of these two filament threads (flux ropes)
and continued for several minutes, implying the interaction between rising FR1 and FR2. Then south ends of FR2 and
FR1 brightened in tandem at 11:54:41 UT and 11:54:29 UT, respectively (Figures 3(a2)-3(a3)). At 11:54:41 UT,
brightening at the south cross site of FR1 and FR2 was detected as well. The two flux ropes then were tracked completely
by the brightening material, and FR2 began to moved upward as well (see panel (a5)). In panel (a6), we can see that
FR2 showed obvious twisted threads while FR1 had erupted outwards. At 11:56:08 UT, the writhed structure of FR2
appeared (see panel (a7)). As shown in Figure 2(c), at 11:24 UT, the maximum $\mid$$T_{w}$$\mid$ of FR2 was beyond
1.75, which is the threshold value of kink instability (T{\"o}r{\"o}k et al. 2004). Combined with the observations
of writhed structure, the high $T_{w}$ of FR2 implied the occurrence of kink instability. In 94 {\AA} channel, the
two flux ropes were also observed clearly at 11:55:23 UT (panel (b)).

Through checking the HMI data, we notice that the double-decker flux rope configuration was lying above a semicircular
PIL (see Fig. 4). It is shown that the south ends of the two flux ropes were rooted in positive magnetic fields to the
southwest of the PIL, and their north ends in a negative-polarity patch to the northeast of the PIL. Before the onset
of the X9.3 flare, this negative magnetic patch kept moving northwestward along the semicircular PIL and successively
sheared with the adjacent positive fields  (see panels (a1)-(a3) and the corresponding
animation). Meanwhile, the HMI continuum intensitygrams reveal that this negative patch exhibited a counterclockwise
rotation motion (panel (b1)). The shearing motion and rotation of this negative magnetic patch can be seen clearly
in the velocity map of panel (c). As a result, the twist number of the two flux ropes rooted in this negative
patch could gradually increase, and eventually approach or exceed the threshold value of kink instability, leading
to the onset of this large flare. Furthermore, when the flare occurred, a white-light flare can be identified in the
HMI continuum intensitygram (panel (b2)), and the LOS magnetogram exhibited magnetic transients as well (panel (a2)).

\begin{figure*}
\centering
\includegraphics [width=0.96\textwidth]{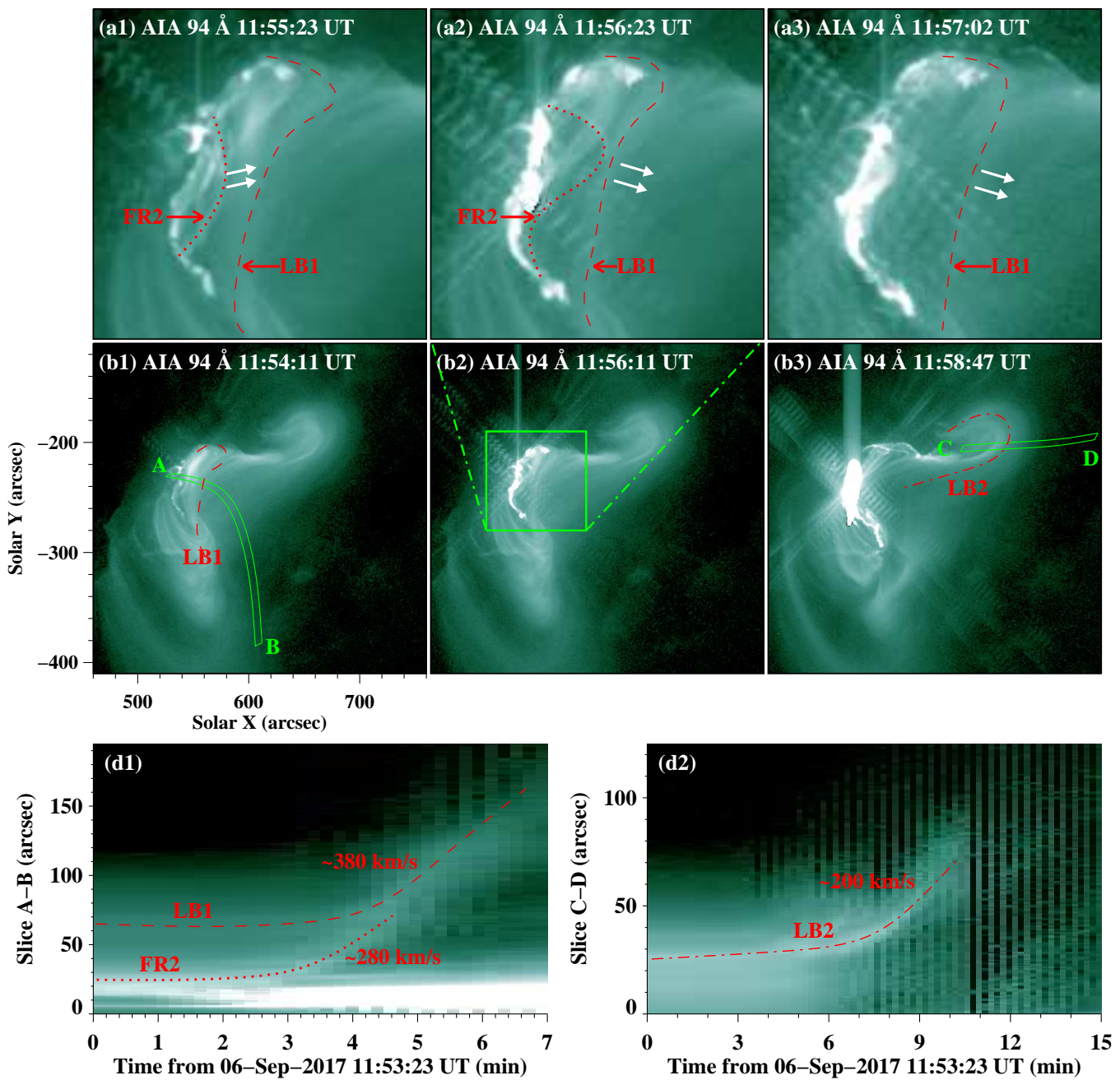}
\caption{
Dynamic evolutions of the complex system consisting of multiple flux ropes and twisted loop bundles during
the X9.3 flare.
Panels (a1)-(a3): sequence of AIA 94 {\AA} images showing the interaction between FR2 and the first twisted loop
bundles (LB1). The red curves delineate their main axes. The white solid arrows in panel (a1) mark the rising
directions of FR2. The white arrows in panels (a2)-(a3) denote the moving direction of LB1 after the interaction
with FR2.
Panels (b1)-(b3): 94 {\AA} images with a larger FOV exhibiting LB1 and the second twisted loop bundles (LB2).
The green square in panel (b2) outlines the FOV of panels (a1)-(a3).
Panels (d1)-(d2): time-space plots along arc-sector domains ``A-B'' of panel (b1) and ``C-D'' of panel (b3) in
94 {\AA} channel. The red lines approximate trajectories of FR2, LB1, and LB2.
The full temporal evolution of the 94 {\AA} images is available as a movie (3.mpg) in the online edition.
}
\label{fig5}
\end{figure*}

\begin{figure*}
\centering
\includegraphics [width=0.96\textwidth]{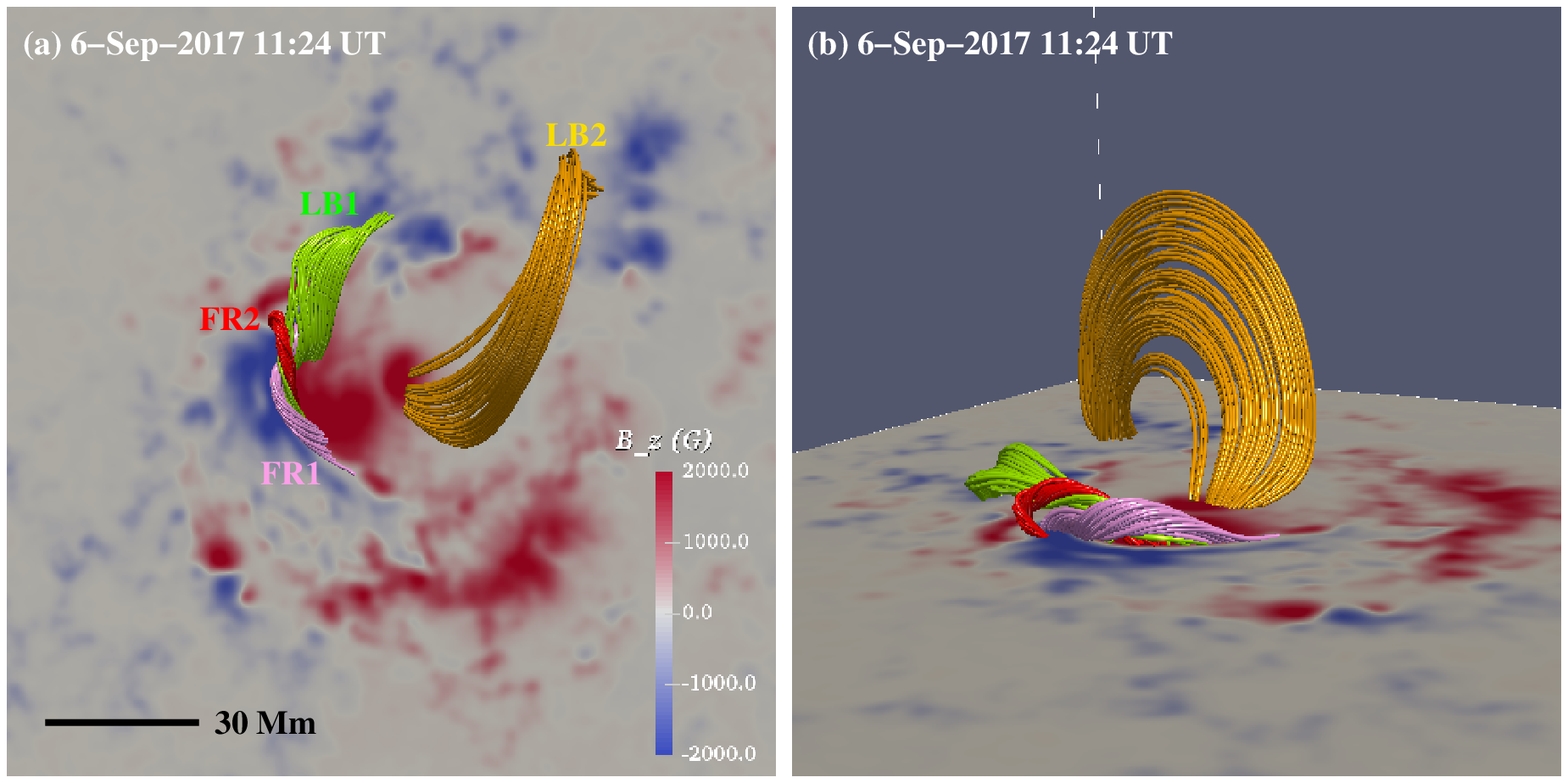}
\caption{
Extrapolated 3D NLFFF structures corresponding to FR1, FR2, LB1, and LB2 at 11:24 UT on 2017 September 6.
Panel (a) shows these structures from a top view, and panel (b) shows them from a side view.
The FOV of this figure is similar to that of Figs. 1(c)-(d).
}
\label{fig6}
\end{figure*}

During the X9.3 flare, we identify a total of two flux ropes and some twisted loop bundles in the flaring
region (see Fig. 5 and the corresponding animation). In 94 {\AA} images of panels (a1)-(a3), we show the
interaction between the kink-unstable FR2 described in Figs. 2 and 3, and the nearby loop bundles (LB1). They are
outlined respectively with red dotted and dashed curves. At 11:55:23 UT, FR2 was rising in the northwest direction,
approaching the adjacent LB1. Around 11:56:23 UT, FR2 and LB1 interacted with each other in their middle parts
(panel (a2)). Then LB1 began to rise up rapidly (panel (a3)). In panels (b1)-(b3), we show LB1 and another
set of loop bundles (LB2) with a larger scale. The green arc-sector domain ``A-B'' in panel (b1) is approximately
along the erupting directions of FR2 and LB1, and the domain ``C-D'' in panel (b3) is in the direction of LB2.
Along the two arc-sector domains, we make two time-space plots and show them in panels (d1) and (d2), respectively.
Panel (d1) shows that FR2 began to rise upward at about 11:55:30 UT, accompanied by the brightening at its base.
Around 11:57:30 UT, FR2 approached LB1 with a projected velocity of $\sim$280 km s$^{-1}$, and then FR2 started
to erupt with a speed of $\sim$380 km s$^{-1}$. As shown in panel (d2), LB2 was disturbed around 12:00 UT and
then erupted with a projected velocity of $\sim$200 km s$^{-1}$. After the successive eruptions of
multiple flux ropes and twisted loop bundles, the X9.3 flare reached its peak at 12:02 UT.

In order to verify these structures illuminated in EUV channels and study their magnetic topologies,
we reconstruct 3D magnetic field above the AR and select a larger FOV involving all the structures mentioned
above to show in Fig. 6. Similar to the analysis of Fig. 2, after calculating the $T_{w}$ of the
reconstructed field, we plot the field lines across the photosphere where the $\mid$$T_{w}$$\mid$ $\geqq$ 1.0
near the PIL in the AR core region and then get two magnetic flux ropes: the pink FR1 and the red FR2.
Resetting the filter value of $\mid$$T_{w}$$\mid$ as 0.5, we obtain a set of green twisted loop bundles beside
the FR2, which corresponds to the LB1. Around the main sunspot, we track the field lines and get another set of
orange loop bundles (LB2) with a smaller $\mid$$T_{w}$$\mid$. Panels (a) and (b) exhibit these structures from
the top view and side view, respectively. It is worthy noting that, due to the projection effect, these structures
located in the southwest region of the solar surface would present different shapes in the remotely-sensed images
compared with that seen from the top view in the reconstructed field by NLFFF modeling.

\subsection{The formation and evolution of the complex magnetic fields of AR 12673}

\begin{figure*}
\centering
\includegraphics [width=.96\textwidth]{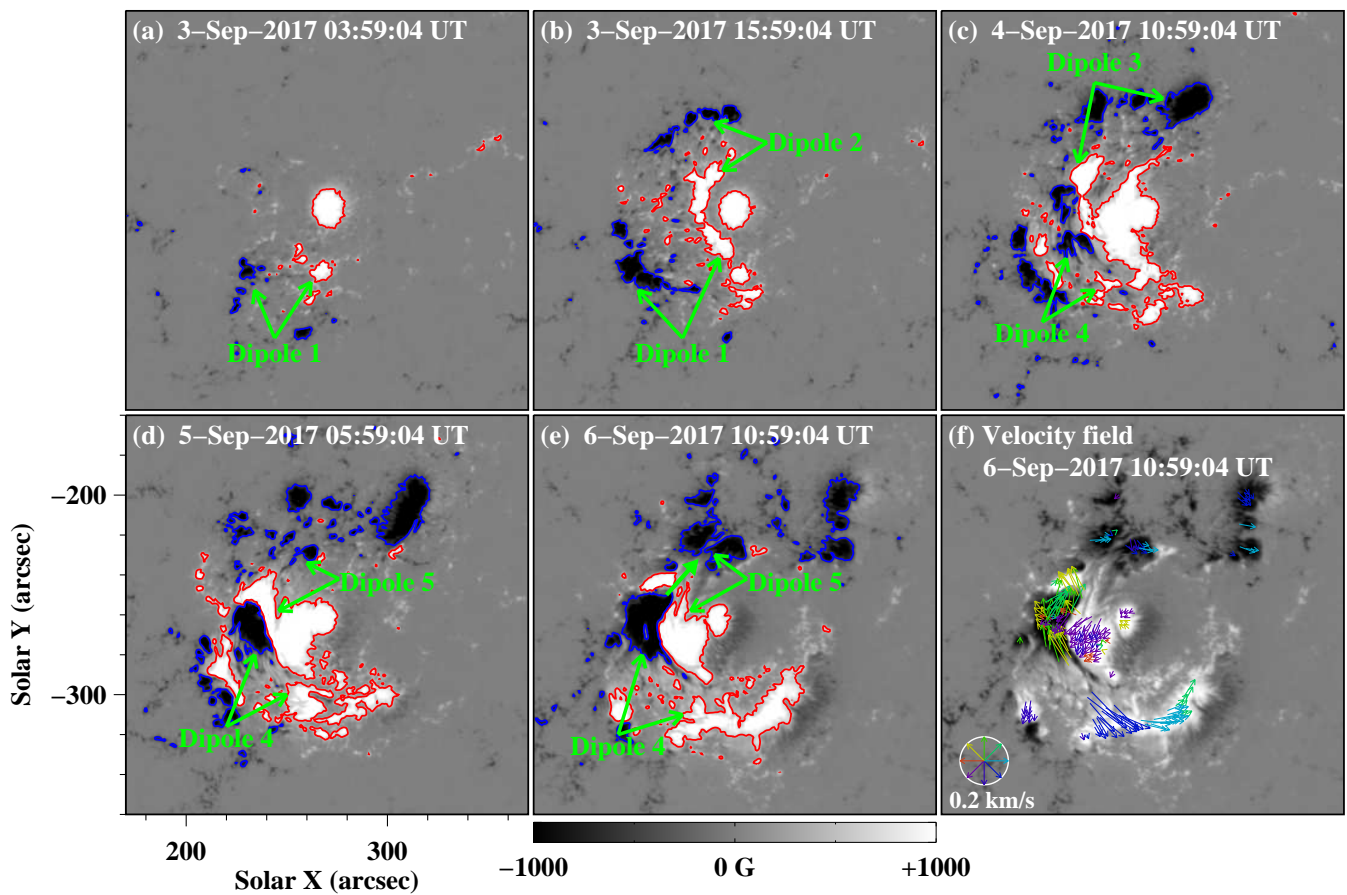}
\caption{
Sequence of HMI LOS magnetograms displaying the evolution of magnetic fields in AR 12673 from
September 3 to September 6. The red and blue curves are contours of the LOS magnetograms at
+500 and -500 G, respectively. Five newly emerging dipoles are marked by green arrows. The green
solid arrow in panel (e) denotes the moving direction of the negative patch of ``Dipole 4''.
Panel (f) shows the horizontal photospheric velocity field (colored arrows) derived from the HMI
LOS magnetograms computed by the DAVE method. Arrows are only superimposed at locations where
the absolute value of magnetic field is greater than 1000 G and illustrate the horizontal flows
around 10:59:04 UT on September 6. The FOV of these panels is the same as that of Figs. 1(c)-(d).
The temporal evolution of the HMI intensitygrams and HMI LOS magnetograms is available as a movie
(4.mpg) online.
}
\label{fig7}
\end{figure*}

Successive eruptions of multiple flux ropes and twisted loop bundles during the X9.3 flare implies the
existence of a complex magnetic system in the AR core region. To investigate the formation and evolution of
such a complicated magnetic system, we analyze the HMI observations in AR 12673 (see Fig. 7 and the corresponding
animation). At the initial stage of the AR evolution, there was simply one main sunspot with positive polarity
in the AR center. Then on September 3, one pair of dipolar fields emerged to the southeast of the main sunspot
(see the ``Dipole 1'' in Fig. 5(a)), followed by the emergence of another dipolar region to the northeast of
the main sunspot (see the ``Dipole 2'' in panel (b)). After this, the negative and positive patches of ``Dipole 1''
separated in the east-west direction, as well as the patches of ``Dipole 2''. Due to the existence of the main
sunspot with strong fields, the positive patches of these two dipoles were blocked (Yang et al. 2017). Thus, an
elongated positive region was formed on the east side of the main sunspot (see panel (b)). Between the negative
and positive patches of these two dipoles, a semicircular channel was formed. Then on September 4, two pairs of
dipolar fields newly emerged within this semicircular channel (see the ``Dipole 3'' and ``Dipole 4'' in panel (c)),
and their patches with opposite polarities separated along the north-south axis of the channel. Here we
speculate that magnetic reconnection took place between ``Dipole 1'' and ``Dipole 4'', forming the FR1 and FR2
whose north ends were rooted in the negative patch of ``Dipole 4'' and south ends rooted in the positive patch of
``Dipole 1''. A similar process could have occurred between ``Dipole 2'' and ``Dipole 3'' and formed LB2.

The positive patch of ``Dipole 3'' moved southeastward while the negative one of ``Dipole 4'' moved toward the
northwest, which then collided with each other eventually (panel (d)). Meanwhile, a new bipolar region ``Dipole 5''
emerged to the west of the positive patch of ``Dipole 3'' around September 5. We suggest that part of the loops
connecting the opposite patches of ``Dipole 5'' would eventually evolve to LB1 due to the rotation of their
footpoints. On September 6, the negative patch of ``Dipole 4'' quickly intruded into the northwest positive region
(see the green solid arrow in panel (e)), which contributed to enhancement of $T_{w}$ of FR1 and FR2 and led to the
onset of the X9.3 flare as mentioned in Fig. 4. The velocity field in panel (f) derived from the DAVE method shows
that the velocity of the patch's motion reached up to $\sim$0.35 km s$^{-1}$ at 10:59:04 UT, before the X9.3 flare's
onset.

\subsection{The X8.2 flare on September 10}

\begin{figure*}
\centering
\includegraphics [width=0.96\textwidth]{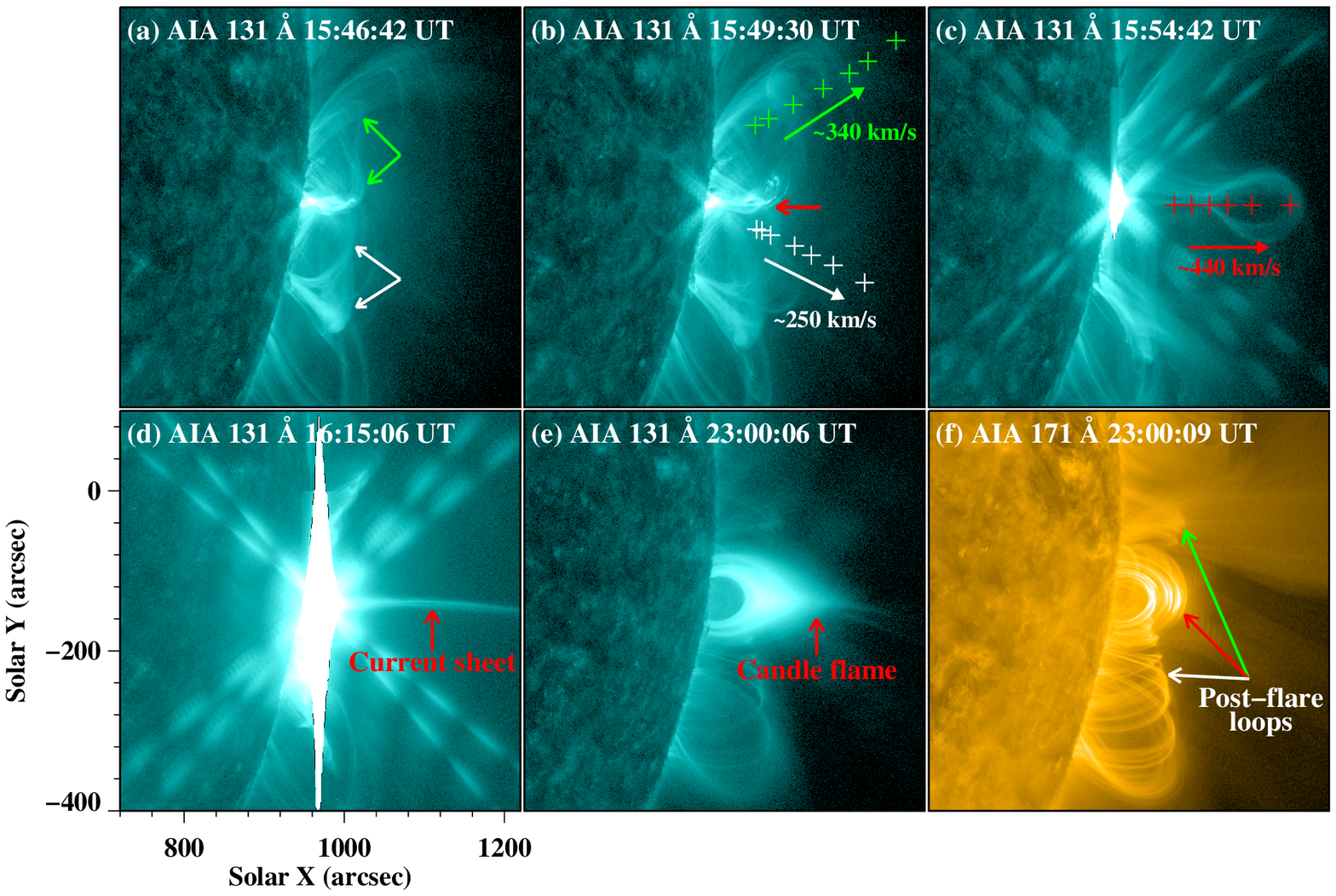}
\caption{
Sequence of AIA 131 {\AA} and 171 {\AA} images showing the evolution of another X8.2 flare in AR 12673
on 2017 September 10. Multiple ejecta components and their counterparts before eruption are denoted by the
pluses and arrows with different colors. The red arrows in panels (d) and (e) mark the current sheet and
candle-flame-shaped structure in the decay phase of the flare. The AIA 171 {\AA} image in panel (f) displays
three groups of post-flare loops with different orientations.
An animation (5.mpg) of the 171 {\AA} and the 131 {\AA} images is available online.
}
\label{fig8}
\end{figure*}

On September 10, an X8.2 flare took place in AR 12673, which was near the west solar limb. In this event,
we also detect multiple ejecta components erupting consecutively and their counterparts (twisted structures)
before the flare peaked at 16:06 UT (see Fig. 8 and the corresponding animation). These ejected structures
were observed clearly in the 131 {\AA} channel and separately denoted by the white, green, and red arrows
in panels (a)-(b)). At about 15:35 UT, the emission at the north end of the twisted structure denoted by
white arrows was enhanced. The brightening propagated southwards and traced out the whole structure, which kept
rising slowly during the following ten minutes. Around 15:46 UT, another twisted structure marked by green arrows
was illuminated. Then the two structures erupted outward with projected velocities of $\sim$250 km s$^{-1}$ and
$\sim$340 km s$^{-1}$, respectively (see panel (b)). Meanwhile, one tear-drop-shaped structure appeared and
rapidly moved outward with a velocity of $\sim$440 km s$^{-1}$ (see panel (c)). After the eruption of the
tear-drop-shaped structure, an evident current sheet was formed in the core region and lasted for several hours
(panel (d)). Moreover, the post-flare candle-flame-shaped structure (Guidoni et al. 2015; Gou et al. 2015, 2016)
was detected clearly in the 131 {\AA} channel (see panel (e)). In the 171 {\AA} channel, one can see three groups
of post-flare loops with different orientations in the flaring region (see the arrows with different colors in
panel (f)). These post-flare loops probably resulted from the eruptions of the multiple twisted structures
mentioned above.

\section{Conclusions and discussion}

Employing the \emph{SDO} observations, we investigate the two largest flares of Solar Cycle 24 occurring
in AR 12673 and the evolution of the AR magnetic fields. On 2017 September 6, the largest flare of Solar
Cycle 24 took place with its peak intensity reaching X9.3. Aided by NLFFF modeling, we identify
a double-decker flux rope configuration above the PIL in the AR core region. The north ends of these two flux
ropes were rooted at a negative magnetic patch, which began to move along the PIL and kept shearing with
adjacent positive fields before the X9.3 flare on September 6. The strong shearing motion as well as a
continuous rotation contributed together to the destabilization of the two magnetic flux ropes. Then the
upper flux rope erupted upward due to the kink-instability and led to the successive eruptions of another
two sets of twisted loop bundles beside the flux ropes within five minutes like a chain reaction. Similarly,
during another X8.2 flare occurring on September 10, we also detected the successive eruptions of multiple
ejecta components. The evolution of the AR magnetic fields shows that five dipoles emerged successively at
the east of the main sunspot. The interactions between these dipoles took place continuously, accompanied by
magnetic flux cancellations and strong shearing motions.

Flux ropes have been thought to be closely connected with CMEs and solar flares (Lin \& Forbes 2000; Fan 2005;
Liu 2013). Amari et al. (2000) proposed a model to approach the theory of CMEs and two-ribbon flares, in which
twisted flux ropes play a crucial role. It was shown that the modeled magnetic configuration could not stay in
equilibrium, and a considerable amount of magnetic energy was released during the eruption of the flux rope.
Employing high-resolution observations from space platforms, Zhang et al. (2015) detected 1354 flux rope proxies
over the solar disk from 2013 January to 2013 December. Hou et al. (2016) further implied the existence of
multiple flux ropes during the evolution of AR 11897. The classical scenario assumes a single flux rope
for each eruption, but it is easy to imagine multiple flux ropes if the AR is complex and has extended curved
PIL (Liu et al. 2009; Liu et al. 2012; Shen et al. 2013; Awasthi et al. 2018). T{\"o}r{\"o}k et al. (2011)
presented a 3D MHD simulation to investigate three consecutive filament eruptions. They considered a configuration
that contains two coronal flux ropes located within a pseudo-streamer and one rope located next to it. It is found
that a sequence of eruptions was initiated by the eruption of the flux rope next to the streamer. The expansion
of this rope resulted in two successive reconnection events, each of which triggered the eruption of a flux rope
by reducing the overlying stabilizing flux. In the observational domain, Shen et al. (2012) reported the
simultaneous occurrence of a partial and a full filament eruption in two neighboring source regions.
Cheng et al. (2013) investigated successive eruptions of two flux ropes with an interval of several hours.
In the present work, we identify a double-decker flux rope configuration above the PIL in the AR core
region. The two flux ropes (FR1 and FR2) erupted at the onset of the X9.3 flare due to the shearing motion
and rotation of the negative magnetic patch where the ropes were rooted. Then another two sets of twisted
loop bundles (LB1 and LB2) beside these ropes were disturbed and successively erupted within five minutes like
a chain reaction. The results from NLFFF modeling show that the $\mid$$T_{w}$$\mid$ of FR1 and FR2 are
beyond 1.0 and the $\mid$$T_{w}$$\mid$ of LB1 is beyond 0.5. If we take a lower standard for defining a magnetic
flux rope (e.g., Chintzoglou et al. 2015, who consider a half turn to be sufficient), then LB1 could be regarded
as the third flux rope in this event. Therefore, we propose that the eruptions of a multi-flux-rope system
rapidly released enormous magnetic energy and led to the X9.3 flare on September 6, the largest flare in Solar
Cycle 24. Similar phenomenon was also observed during another X8.2 flare occurring in the same AR several days
later.

In recent years, the concept of double-decker filament (flux rope) was proposed by Liu et al. (2012)
to explain two vertically separated filaments (flux ropes) over the same PIL. The complex configuration of
double-decker flux rope was observed and modeled to exist prior to solar eruptive events (Cheng et al. 2014;
Kliem et al. 2014). Extrapolated NLFFF structures in this work reveal that before the onset of the X9.3 flare,
two magnetic flux ropes were located separated vertically above the PIL in the AR core region, forming a typical
double-decker flux rope configuration. At the onset of the X9.3 flare, the strong shearing motion and rotation of
the north ends of the two magnetic flux ropes contributed to their destabilization (Kliem et al. 2004;
Srivastava et al. 2010; T{\"o}r{\"o}k et al. 2013; Yan et al. 2018b). The brightening at the cross sites of
these two ropes observed in EUV wavelength indicated the interaction (magnetic reconnection) occurring between
the two flux ropes during their slow-rise phase. The subsequent AIA observations revealed that the lower rope
lost its stability first and erupted outwards while the upper flux rope kept rising upward. Then the upper magnetic
flux rope writhed into a sigmoid shape. The calculation of twist number based on the NLFFF results shows that the
maximum $\mid$$T_{w}$$\mid$ of the two flux ropes were all beyond 1.75 half an hour before the onset of the flare.
It is worth noting that the exact value of twist required of the kink instability depends on various factors
such as the flux rope geometry and the surrounding magnetic fields. T{\"o}r{\"o}k et al. (2004) proposed that the
threshold of instability increases with rising aspect ratio and the number is 1.75 (3.5 $\pi$) at a loop aspect
ratio $R/a \approx 5$, which corresponds to a rather fat flux rope. Although the double-decker flux ropes in the
present work may have a different aspect ratio, we approximate the threshold value of kink instability to 1.75 here.
The facts that the twist of the flux rope is beyond the threshold value of kink instability and its conversion into
the writhe support the occurrence of the kink instability during the eruption of the upper flux rope. The
eruption of a kink-unstable flux rope during this event has also been investigated by Yang et al. (2017).

The existence of multiple flux ropes and twisted loop bundles during the two X-class flares reported in the
present paper implies a complicated magnetic system in AR 12673. Examining the evolution of the magnetic fields
in the AR core region, we notice that significant flux emergence occurred in this region (Sun \& Norton 2017).
Five dipoles emerged successively at the east of the main sunspot. The negative and positive patches of the first
two dipoles separated along the east-west direction. However, the patches of the latter two dipoles separated along
the north-south direction, perpendicular to the former one. The cross separation of these dipole patches with
opposite polarities led to the continuous interactions between different dipolar fields, accompanied by magnetic
flux cancellations at some places (Toriumi et al. 2013; Louis et al. 2015; Yang et al. 2017). Strong
shearing motions between the patches with opposite polarities accumulated dramatic free energy (Shimizu et al. 2014;
Toriumi \& Takasao 2017) and could result in magnetic reconnection (Moore et al. 2001), which would lead to the
formation of flux ropes (Xue et al. 2017). The rotation of the associated magnetic patch also contributes to the
magnetic flux rope buildup. In Sect. 3.3, we speculated on the detailed process concerning the formations of
these flux ropes and twisted loop bundles. It is worth mentioning that during the impulsive phase of the X9.3 flare,
a white-light signal was detected as well as anomalous magnetic transient near the PIL (see the animation
corresponding to Fig. 4), indicating a violent release of energy (Hudson et al. 1992; Song et al. 2018).
We propose that in AR 12673 significant flux emergence and successive cross-separations between the patches of
different newly emerging dipoles resulted in the formation of multiple flux ropes and twisted loop bundles
in the same AR and the storage of dramatic magnetic energy.

\begin{acknowledgements}
The authors are grateful to the anonymous referee for valuable suggestions.
The data are used courtesy of the \emph{SDO} and \emph{GOES} science teams. \emph{SDO} is a mission of NASA's Living
With a Star Program. This work is supported by the National Natural Science Foundations of China (11533008, 11790304,
11773039, 11673035, 11673034, 11873059, and 11790300), the Youth Innovation Promotion Association of CAS (2017078 and
2014043), and Key Programs of the Chinese Academy of Sciences (QYZDJ-SSW-SLH050).
\end{acknowledgements}

%
%

\clearpage

\end{document}